\documentclass[prl,aps,
twocolumn,
nofootinbib,
floatfix,
superscriptaddress,10pt]{revtex4-2}

\pdfoutput=1

\usepackage[dvipsnames]{xcolor}
\definecolor{mathematicablue}{rgb}{0.368417, 0.506779, 0.709798} 
\definecolor{darkblue}{HTML}{334C70}

\usepackage{amsmath,amssymb,mathrsfs,amsthm,tikz,shuffle,paralist,accents}

\usepackage[hidelinks]{hyperref}
\hypersetup{
    pdfencoding=unicode,
	colorlinks=true,
	urlcolor=Maroon,
    allcolors=Maroon,
	linkcolor=OliveGreen,
	citecolor=Maroon,
	pdftitle={Minimal Cuts and Genealogical Constraints},
	pdfauthor={},
	pdfdisplaydoctitle=true,
	pdfstartview=FitH,
	linktocpage=true
}

\usepackage{graphicx}
\usepackage{soul}

\usetikzlibrary{shapes.misc}
\usetikzlibrary{cd}
\usetikzlibrary{snakes}
\usetikzlibrary{calc}
\usetikzlibrary{fadings}
\usetikzlibrary{decorations.markings}
\usetikzlibrary{positioning}

\tikzset{
quark/.style={postaction={decorate},
  decoration={markings,mark=at position .5 with {\arrow[#1]{latex}}}},
gluon/.style={decorate,
 decoration={coil, aspect=1, amplitude=2pt, segment length=6pt,  pre length=0cm, post length=0cm}},
dphoton/.style={decorate, decoration={snake, segment length=8pt, amplitude=1.8pt,  pre length=0cm, post length=0cm}}
}

\usepackage[all]{xy}

\newenvironment{definition}%
  {\list{}{\leftmargin=0.3cm\rightmargin=0.3cm}\item[]}%
  {\endlist}

\tikzset{
    ncbar angle/.initial=90,
    ncbar/.style={
        to path=(\tikztostart)
        -- ($(\tikztostart)!#1!\pgfkeysvalueof{/tikz/ncbar angle}:(\tikztotarget)$)
        -- ($(\tikztotarget)!($(\tikztostart)!#1!\pgfkeysvalueof{/tikz/ncbar angle}:(\tikztotarget)$)!\pgfkeysvalueof{/tikz/ncbar angle}:(\tikztostart)$)
        -- (\tikztotarget)
    },
    ncbar/.default=0.5cm,
}
\tikzset{round left paren/.style={ncbar=0.5cm,out=115,in=-115}}
\tikzset{round right paren/.style={ncbar=0.5cm,out=60,in=-60}}

\def\F{\mathcal{F}}
\def\U{\mathcal{U}}
\def\d{\mathrm{d}}
\def\disc{\Delta}

\def\I{I}

\begin{document}

\title{Minimal Cuts and Genealogical Constraints on Feynman Integrals}

\author{Holmfridur~S.~Hannesdottir}
\affiliation{Institute for Advanced Study, Einstein Drive, Princeton, NJ 08540, USA}
\author{Luke~Lippstreu}
\affiliation{Higgs Centre for Theoretical Physics, School of Physics and Astronomy,
The University of Edinburgh, Edinburgh EH9 3FD, Scotland, UK}
\author{Andrew~J.~McLeod}
\affiliation{Higgs Centre for Theoretical Physics, School of Physics and Astronomy,
The University of Edinburgh, Edinburgh EH9 3FD, Scotland, UK}
\author{Maria~Polackova}
\affiliation{Higgs Centre for Theoretical Physics, School of Physics and Astronomy,
The University of Edinburgh, Edinburgh EH9 3FD, Scotland, UK}

\begin{abstract} \noindent
We introduce an efficient method for deriving hierarchical constraints on the discontinuities of individual Feynman integrals. This method can be applied at any loop order and particle multiplicity, and to any configuration of massive or massless virtual particles. The resulting constraints hold to all orders in dimensional regularization, and complement the extended Steinmann relations---which restrict adjacent sequential discontinuities---by disallowing ordered pairs of discontinuities from appearing even when separated by (any number of) other discontinuities. We focus on a preferred class of hierarchical constraints, which we refer to as \emph{genealogical constraints}, that govern what singularities can follow from certain \emph{minimal cuts} that act as the primogenitors of the discontinuities that appear in Feynman integrals. While deriving the full set of hierarchical constraints on a given Feynman integral generally requires identifying all solutions to the (blown up) Landau equations, these genealogical constraints can be worked out with only minimal information about what singularities may appear.  We illustrate the power of this new method in examples at one, two, and three loops, and provide evidence that genealogical constraints restrict the analytic structure of Feynman integrals significantly more than the extended Steinmann relations.
\end{abstract} 

\ \vspace{.1cm}
\maketitle

\noindent {\bf Introduction}
\vspace{.1cm}

Methods for studying the analytic structure of perturbative scattering amplitudes have been experiencing a renaissance~\cite{Abreu:2017ptx,Abreu:2017enx,Bourjaily:2020wvq,Mizera:2021icv,Hannesdottir:2021kpd,Hannesdottir:2022bmo,Mizera:2022dko,Hannesdottir:2022xki,Berghoff:2022mqu,Britto:2023rig,Fevola:2023kaw,Fevola:2023fzn,Vergu:2023rqz,Helmer:2024wax}. Much of the resurgent interest in this topic has stemmed from the striking analytic features that amplitudes have been observed to exhibit in recent years, ranging from intriguing cluster-algebraic structures to surprising dualities~\cite{Golden:2013xva,Schnetz:2013hqa,Golden:2014xqa,Golden:2014pua,Panzer:2016snt,Drummond:2017ssj,Schnetz:2017bko,Henn:2019mvc,Golden:2019kks,Caron-Huot:2019bsq,Chicherin:2020umh,Dixon:2021tdw,He:2021eec,Abreu:2021smk,Dixon:2022xqh,Henn:2024ngj}. By better understanding how the analytic properties of these amplitudes follow from basic physical principles, it is hoped that new insights might be gleaned into how these features can be incorporated into our best computational methods, thereby helping us evaluate important quantities such as Standard Model amplitudes. 

In this letter, we make progress towards this goal by elucidating an important aspect of the analytic structure exhibited by individual Feynman integrals. We do this by introducing a practical method for working out the implications of the \emph{hierarchical principle}, which traces its origins back to the sixties~\cite{pham,Landshoff1966,boyling1968homological,Hannesdottir:2022xki,Berghoff:2022mqu}. While it has long been known that this principle places strong constraints on the discontinuities of Feynman integrals, deriving these constraints in practice generally requires implementing a number of algebraic blowups to uncover the full set of singularities that can appear in a given integral. However---drawing on recent advances in both the understanding of Feynman-parameter space cuts~\cite{Berghoff:2022mqu,Britto:2023rig} and the use of topological invariants for probing where Feynman integrals can become singular~\cite{Fevola:2023fzn,Fevola:2023kaw}---we show that this difficulty can be sidestepped for a large class of hierarchical constraints. Namely, we present an efficient method for identifying singularities that can no longer be accessed after one computes one or more discontinuities of one's Feynman integral, by tracking how these discontinuities modify the integration contour in Feynman parameter space; all of this can be done without working out any explicit blowups. The resulting constraints, which we call \emph{genealogical} as they describe which singularities cannot descend from special classes of \emph{minimal cuts}, can be derived for integrals involving any configuration of massive or massless particles, and hold to all orders in dimensional regularization.

\vspace{.1cm}
\noindent {\bf Feynman Integrals and the Hierarchical Principle} \!\!\!\!\!\!\!
\vspace{.1cm}

The constraints we study in this letter apply to individual scalar Feynman integrals, namely integrals that take the form
\begin{equation}
    \label{eq:Feynman_integral}
    I(p_i) = \frac{(-1)^E}{(i \pi)^{L D/2}} \int \frac{\d^D k_1\cdots d^D k_L}{A_1 \cdots A_E} \, ,
\end{equation}
where $L$ is the number of loops, $E$ is the number of internal propagators, and $D = D_0- 2\epsilon$ denotes the spacetime dimension (where $D_0$ is an integer). The propagators $A_i$ are quadratic polynomials in the external momenta $p_i$, the loop momenta $k_i$, and any internal masses $m_i$.

It is often convenient to recast~\eqref{eq:Feynman_integral} into a parametric form by introducing Feynman parameters and integrating out the loop momenta, resulting in
\begin{equation}
\label{eq:Feynman_integral_feyn_param}
    I(p_i) = \Gamma(d) \lim_{\varepsilon \to 0^+} \int_0^\infty \frac{ \d \alpha_1 \cdots \d \alpha_E}{\text{GL}(1)} \frac{\U^{d-D/2}}{(-\F- i \varepsilon)^{d}} \, ,
\end{equation}
where $\U$ and $\F$ are polynomials of the Feynman parameters $\alpha_i$, which can be defined in terms of spanning trees of the corresponding Feynman graph (see for instance~\cite{Weinzierl:2022eaz}), and $d=E-L D/2$. We can use the $\text{GL}(1)$ covariance to fix, for example, $\sum_e \alpha_e = 1$. In this form, the integral is manifestly a function of Lorentz-invariant combinations of the external kinematics. 
Although we have left the dependence on internal and external masses implicit, we note that we allow these masses to take any value, including zero.

Much of the physical wisdom incorporated into Feynman integrals manifests itself through their analytic structure. Important aspects of this structure are captured by the Landau equations~\cite{Bjorken:1959fd,Landau:1959fi,10.1143/PTP.22.128,Fevola:2023kaw, Fevola:2023fzn}, which identify where Feynman integrals can become singular, and the \emph{Steinmann relations}~\cite{Steinmann,Steinmann2,Cahill:1973qp,Caron-Huot:2016owq,Bourjaily:2020wvq}, which tell us that double discontinuities with respect to partially-overlapping momentum channels must vanish. 
Another intriguing facet is encapsulated by the \emph{hierarchical principle}~\cite{pham,boyling1968homological,Landshoff1966,Berghoff:2022mqu,Hannesdottir:2022xki}. In its original formulation, this principle asserts that once a discontinuity with respect to a given singularity has been computed, we can no longer access singularities that require taking any of the propagators $A_i$ that were involved in pinching the integration contour back off shell. 

A version of the hierarchical principle has also been proven in Feynman parameter space~\cite{Berghoff:2022mqu}, from which it follows that the discontinuities of Feynman integrals can be expressed in terms of integrals of the form~\eqref{eq:Feynman_integral_feyn_param} in which the integration contour has been modified. More specifically, the discontinuity
\begin{equation}
    \Delta_{\lambda} \I = \I^\circlearrowleft_{\lambda} - \I \, ,
    \label{eq:disc_def}
\end{equation}
which corresponds to the difference between the value of the integral before and after being analytically continued around the branch point at $\lambda=0$, can be written as
\begin{equation}
    \Delta_{\lambda} \I
    = c
    \int_{h_\lambda} \d \I \,,
\end{equation}
where $c$ is a number, $\d I$ is the same integrand that appeared in~\eqref{eq:Feynman_integral_feyn_param}, and $h_\lambda$ is a new integration contour. 

In what follows, we will not need the explicit form of the contours $h_\lambda$. Rather, we only use the weaker result that this contour is contained within a small ball around the corresponding pinched surface. Already, this implies that the new contour $h_\lambda$ must forget about the original $\alpha_i=0$ integration boundary whenever the singularity at $\lambda=0$ arises from a pinch of the $\alpha_i$ contour. It follows that the corresponding endpoint singularities---which occur when the integrand becomes singular at the $\alpha_i = 0$ boundary of integration, rather than due to the pinching of the $\alpha_i$ integration contour---can no longer be accessed.

This phenomenon has already be seen in numerous examples in the literature~\cite{Britto:2023rig,Berghoff:2022mqu}. For instance, the integration contours that result from computing different discontinuities of the massive triangle integral were presented in~\cite{Britto:2023rig}. There, it can be seen that discontinuities with respect to internal masses---which correspond to singularities in which just a single Feynman parameter contour is pinched---give rise to contours that are bounded by the two remaining $\alpha_{i} = 0$ lines and $\F=0$. Conversely, discontinuities with respect to threshold singularities---in which two Feynman parameter contours are pinched---result in integration contours that are bounded by just a single $\alpha_i=0$ line and $\F=0$. Finally, the discontinuity associated with the maximal cut forgets about all of the $\alpha_i=0$ boundaries and results in an integration contour that is just bounded by $\F=0$.

\vspace{.1cm}
\noindent {\bf The Two-Mass Easy Box}
\vspace{.1cm}

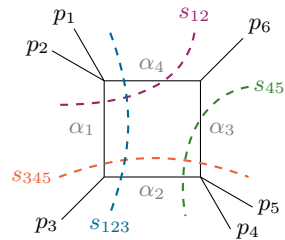
\begin{figure}
    \centering
    \begin{tikzpicture}[scale=.8]
        \coordinate (a) at (-.8,.8);
        \coordinate (b) at (.8,.8);
        \coordinate (c) at (.8,-.8);
        \coordinate (d) at (-.8,-.8);
        \draw[] (a) to (b);
        \draw[] (b) to (c);
        \draw[] (c) to (d);
        \draw[] (d) to (a);
        \draw[] (a) --++ (150:1);
        \draw[] (a) --++ (120:1);
        \draw[] (b) --++ (45:1);
        \draw[] (c) --++ (-30:1);
        \draw[] (c) --++ (-60:1);
        \draw[] (d) --++ (-135:1);
        
        \node[] at (-1.45,1.85) {$p_1$};
        \node[] at (-1.9,1.35) {$p_2$};
        \node[] at (-1.75,-1.6) {$p_3$};
        \node[] at (1.6,-1.7) {$p_4$};
        \node[] at (1.95,-1.3) {$p_5$};
        \node[] at (1.8,1.7) {$p_6$};

        \node at ($(a)!0.5!(b)$) [above]{\color{gray} $\alpha_4$};
        \node at ($(a)!0.5!(d)$) [left]{\color{gray} $\alpha_1$};
        \node at ($(b)!0.5!(c)$) [right]{\color{gray} $\alpha_3$};
        \node at ($(c)!0.5!(d)$) [below]{\color{gray} $\alpha_2$};
        
        \draw[RedViolet,thick,dashed] (0.7,1.6) to [out=-100,in=0] (-1.6,.4);
        \draw[OliveGreen,thick,dashed] (1.6,.7) to [out=-170,in=-260] (.55,-1.45);
        \draw[RedOrange,thick,dashed] (-1.55,-.8) to [out=20,in=-200] (1.6,-.8);
        \draw[MidnightBlue,thick,dashed] (-.7,1.6) to [out=-70,in=70] (-.7,-1.4);

        \node[] at (0.7,1.9) {\color{RedViolet} $s_{12}$};        
        \node[] at (-2,-.8) {\color{RedOrange} $s_{345}$};        
        \node[] at (1.95,.7) {\color{OliveGreen} $s_{45}$};        
        \node[] at (-.7,-1.6) {\color{MidnightBlue} $s_{123}$};        
        
    \end{tikzpicture}
    \caption{The Feynman diagram for the two-mass easy box integral, and its two-particle cuts.}
    \label{fig:two_mass_easy_box}
\end{figure}

The imprint of the hierarchical principle can already be seen at one loop~\cite{pham,Berghoff:2022mqu}. Consider as an example the two-mass easy box (Figure~\ref{fig:two_mass_easy_box}), which evaluates to polylogarithms that draw upon a ten-letter symbol alphabet. In this case, each letter arises from a unique solution to the Landau equations, so we know exactly where the corresponding logarithmic branch points arise in the integral---and in particular, we know which Feynman parameter contours are pinched. This information is given in Table~\ref{tab:two_mass_easy_box_constraints}, where $s_{ij} = (p_i+p_j)^2$ and $s_{ijk} = (p_i + p_j + p_k)^2$.

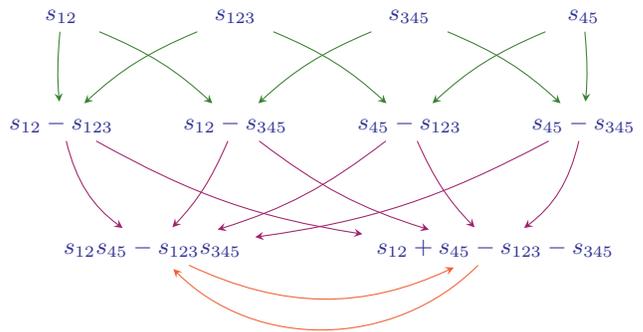
\begin{figure}
  \begin{tikzpicture}[%
    ->,
    shorten >=2pt,
    >=stealth,
    node distance=1cm,
    noname/.style={%
    color=Blue, minimum width=4em, minimum height=.1cm,scale=1}
  ]
    \node[noname] (12) {$s_{12}$};
    \node[noname] (123) [right=of 12] {$s_{123}$};
    \node[noname] (345) [right=of 123] {$s_{345}$};
    \node[noname] (45) [right=of 345] {$s_{45}$};
    \node[noname] (12m123) [below=of 12] {$s_{12} - s_{123}$};
    \node[noname] (12m345) [below=of 123] {$s_{12} - s_{345}$};
    \node[noname] (45m123) [below=of 345] {$s_{45} - s_{123}$};
    \node[noname] (45m345) [below=of 45] {$s_{45} - s_{345}$};
    \node[noname] (12t45m123t345) [node distance=1.2cm and -1cm,below left=of 12m345] {$s_{12} s_{45} - s_{123}s_{345}$};
    \node[noname] (12m123m345p45) [node distance=1.2cm and -1.36cm,below right=of 45m123] {$s_{12} + s_{45} - s_{123} - s_{345}$};
    \path (12) edge [bend right=5pt, color=OliveGreen] node {}(12m123)
          (12) edge [bend left=10pt, color=OliveGreen] node {} (12m345)
          (123) edge [bend right=10pt, color=OliveGreen] node {} (12m123)
          (123) edge [bend left=10pt, color=OliveGreen] node {} (45m123)         
          (345) edge [bend right=10pt, color=OliveGreen] node {} (12m345)
          (345) edge [bend left=10pt, color=OliveGreen] node {} (45m345)
          (45) edge [bend right=10pt, color=OliveGreen] node {} (45m123)
          (45) edge [bend left=5pt, color=OliveGreen] node {} (45m345)
          (12m123) edge [bend right=20pt, color=RedViolet] node {} (12t45m123t345)
          (12m123) edge [bend right=10pt, color=RedViolet] node {} (12m123m345p45)    
          (12m345) edge [bend left=10pt, color=RedViolet] node {} (12t45m123t345)
          (12m345) edge [bend right=10pt, color=RedViolet] node {} (12m123m345p45)
          (45m123) edge [bend left=10pt, color=RedViolet] node {} (12t45m123t345)
          (45m123) edge [bend right=10pt, color=RedViolet] node {} (12m123m345p45)
          (45m345) edge [bend left=10pt, color=RedViolet] node {} (12t45m123t345)
          (45m345) edge [bend left=20pt, color=RedViolet] node {} (12m123m345p45)
          (12t45m123t345) edge [bend right=25pt, color=RedOrange] node {} (12m123m345p45)
          (12m123m345p45) edge [bend left=45pt, color=RedOrange] node {} (12t45m123t345);
  \end{tikzpicture}
  \caption{The allowed sequences of discontinuities in the two-mass easy box. An arrow $a \to b$ indicates that there can exist a singularity at $b=0$ after computing a discontinuity around $a=0$.
  Each singularity should also be thought of as having an arrow that points back to itself.
  }
  \label{fig:two_mass_easy_hierarchicial}
\end{figure}

Given this data, the predictions made by the hierarchical principle can be easily deduced.
Namely, once a Feynman parameter has been involved in a pinch, singularities in which this variable does \emph{not} participate can no longer be accessed. 
This places strong restrictions on the \emph{symbol} of this integral~\cite{Goncharov.A.B.:2009tja,Chen:1977oja,Brown:2009qja,Goncharov:2010jf,Duhr:2011zq}, which encodes all of the sequences of logarithmic discontinuities that it contains. We depict the sequential discontinuities that are allowed by the hierarchical principle in the two-mass easy box---and thus, which \emph{letters} can follow each other in its symbol---in Figure~\ref{fig:two_mass_easy_hierarchicial}. There, we see that only one of the letters in the first row, and one of the letters in the second row, can ever appear in each term of the symbol (although they can appear repeatedly). More generally, once a given symbol term has passed from one row of letters to the next, the letters in the rows above it can no longer appear. One can check that these constraints are indeed obeyed by the integral to all orders in dimensional regularization (using for instance the results of~\cite{Ohnemus:1990za,Bern:1993kr,Duplancic:2000sk} or Appendix A in~\cite{Brandhuber:2005kd}).

\begin{table}[t!]
\centering
\begin{tabular}{|c|c|}
\hline
Singularity & \ Pinched $\alpha_i$ contours \ \\ \hline
$s_{12}$ & $\alpha_1$, $\alpha_4$ \\
$s_{45}$ & $\alpha_2$, $\alpha_3$ \\
$s_{123}$ & $\alpha_2$, $\alpha_4$ \\
$s_{345}$ & $\alpha_1$, $\alpha_3$ \\
$s_{12} {-} s_{123}$ & $\alpha_1$, $\alpha_2$, $\alpha_4$ \\
$s_{12} {-} s_{345}$ & $\alpha_1$, $\alpha_3$, $\alpha_4$ \\
$s_{123} {-} s_{45}$ & $\alpha_2$, $\alpha_3$, $\alpha_4$ \\
$s_{345} {-} s_{45}$ & $\alpha_1$, $\alpha_2$, $\alpha_3$ \\
$s_{12} s_{45} {-} s_{123} s_{345}$ & $\alpha_1$, $\alpha_2$, $\alpha_3$, $\alpha_4$ \\
\ $s_{12} {-} s_{123} {-} s_{345} {+} s_{45}$ \ & $\alpha_1$, $\alpha_2$, $\alpha_3$, $\alpha_4$ \  \\
\hline
\end{tabular}

\caption{The singularity locus and set of Feynman parameter contours that are pinched in each solution to the Landau equations for the two-mass easy box.}
\label{tab:two_mass_easy_box_constraints}
\end{table}

Beyond one loop, the predictions made by the hierarchical principle become harder to discern, as the same kinematic singularity often arises from pinches that occur at different locations along the integration contour. Since no practical method yet exists for finding the complete set of solutions to the Landau equations, this gives rise to two difficulties. First, it means that we cannot generally discern which $\alpha_i=0$ boundaries have been dropped when we compute a discontinuity $\Delta_\lambda$. Second, even if we could decide which boundaries have been dropped, concluding that a second singularity at $\lambda^\prime=0$ can no longer be accessed requires ruling out the possibility that an additional $\lambda^\prime=0$ solution to the (blown up) Landau equations exists even when these boundaries are absent. In the next two sections, we present practical solutions to both of these problems.

\vspace{.1cm}
\noindent {\bf The Euler Characteristic Test}
\vspace{.1cm}

Singularities can only arise in Feynman integrals when the space on which the integration contour is defined degenerates~\cite{pham}. This was made precise in~\cite{Fevola:2023kaw, Fevola:2023fzn}, where it was shown that Feynman integrals only become singular on a kinematic variety $\lambda=0$ if the absolute value of the Euler characteristic $\chi(Y)$ drops in value when $\lambda$ vanishes, where
\begin{equation} \label{eq:feyn_param_space}
Y = \mathbb{C}^{E-1} \Big \backslash \left( \F = 0 \cup \U = 0 \bigcup_{i = 1}^{E} \alpha_i = 0 \right)
\end{equation} 
is the space of Feynman parameters $\alpha_i$ minus the loci where the integrand can become singular, and the boundaries of integration. An implementation of this criterion, which we henceforth refer to as the \emph{Euler characteristic test}, can also be found in~\cite{Fevola:2023fzn,Fevola:2023kaw}. 
This implementation makes use of Theorem 1 in~\cite{Huh_2013}, which states that $|\chi|$ is equal to the number of solutions of the system of equations
\begin{equation}
    \sum_{j} \frac{\mu_j}{f_j} \frac{\partial f_j}{ \partial \alpha_i} 
    + \frac{\nu_i}{\alpha_i} = 0
    \quad \text{ for } i \in \{1,2,\ldots,E\} \,
\label{eq:critical_points}
\end{equation}
for generic $\mu_j$ and $\nu_i$,
where $f_j$ are the factors that appear in the integrand (so $f_1=\F$ and $f_2=\U$ in the Feynman parameter representation).\footnote{If we were to work in the Lee-Pomeransky representation, $Y_\lambda$ would have dimension $E$, and there would just be a single factor $f_1 = \U+\F$.} 
Checking whether this number drops when one sets $\lambda=0$ provides an efficient method for determining whether the Feynman integral~\eqref{eq:Feynman_integral_feyn_param} can become singular on the $\lambda=0$ surface.

We are now in a position to take this Euler characteristic test one step further, by applying it not only to Feynman integrals, but also to their discontinuities. To do so, we leverage the observation highlighted above, that certain $\alpha_i=0$ integration boundaries can be dropped from the integration space whenever we compute a discontinuity. 
In particular, by keeping track of which integration boundaries have been dropped, we can refine the Euler characteristic test by not subtracting the $\alpha_i=0$ surfaces in $Y$ that no longer exist as integration boundaries. 
For instance, if the set of Feynman parameter contours in $\{\alpha_1, \alpha_2, \dots, \alpha_j\}$ are known to be pinched at $\lambda=0$, we can probe whether the discontinuity $\disc_\lambda \I$ can become singular on a different variety $\lambda^\prime=0$ by asking whether $|\chi(Y_{1...j})|$ drops in value when $\lambda^\prime$ vanishes, where
\begin{equation} \label{eq:feyn_param_space_refined}
Y_{1...j} = \mathbb{C}^{E-1} \Big \backslash \left( \F = 0 \cup \U = 0 \bigcup_{i = j+1}^E \alpha_i = 0 \right) \, .
\end{equation}
This corresponds to setting $\nu_i=0$ for $i \in \{1,2,\dots ,j \}$ in~\eqref{eq:critical_points}.\footnote{Strictly speaking, Theorem 1 in~\cite{Huh_2013} assumes that the $\nu_e$ are generic (and in particular, nonzero), so it is not guaranteed to hold in the case of interest to us. However, we do not know of a counterexample, in which the number of the solutions to the critical-point equations~\eqref{eq:critical_points} fails to capture a singularity.} If the value of the Euler characteristic does not drop in this limit, we conclude that $\Delta_{\lambda^\prime} \cdots \Delta_{\lambda} \cdots \I = 0$, where the dots represent any other sequences of discontinuities. In this way, the consequences of the hierarchical principle can be worked out without the need to carry out any blowups.

\vspace{.1cm}
\noindent {\bf Minimal Cuts and Genealogical Constraints}
\vspace{.1cm}

In the last section, we saw how hierarchical predictions can be reliably made once we know which $\alpha_i=0$ boundaries have been dropped. In general, however, we have no way of discerning which Feynman parameters have participated in the relevant pinch when computing a discontinuity. To get around this problem, we now focus on a restricted class of hierarchical constraints whose derivation requires no knowledge about where singularities arise in the space of Feynman parameters. To do so, we introduce the concept of \emph{minimal cuts}, which represent the minimal sets of propagators that must be put on shell in order to resolve a singularity:

\begin{definition}
{\bf Minimal Cut:} Given a Feynman integral and a singular kinematic surface $\lambda(\{s_{i\dots j}\}, \{m_k^2\})=0$, we refer to a set of cut propagators as a \emph{minimal cut} if \emph{(i)} the cut propagators partition the external momenta into the combinations that appear in the Mandelstam variables $\{s_{i\dots j}\}$, \emph{(ii)} each of the internal masses in $\{m_k^2\}$ appears in at least one of the cut propagators (unless this mass has already appeared as one of the Mandelstam variables), and \emph{(iii)} one of the first two properties is no longer satisfied if any of the cut propagators are taken off shell.
\end{definition}

\noindent Note that there may exist more than one minimal cut for a given singularity, and that we can sometimes generate different minimal cuts by rewriting the singularity $\lambda$ in terms of different Mandelstam variables. In general, we will need to consider each of the minimal cuts that can be generated this way.

By identifying what sets of propagators must be put on shell in order to resolve all of the kinematic variables in $\lambda(\{s_{i\dots j}\}, \{m_k^2\})$, minimal cuts identify which Feynman parameters are involved in \emph{any} contour pinch that gives rise to a singularity at $\lambda(\{s_{i\dots j}\}, \{m_k^2\})=0$ in the Feynman-parameter representation~\eqref{eq:Feynman_integral_feyn_param}. We emphasize that if the complete solution set to the Landau equations were known, we would not need to use minimal cuts. Nevertheless, when combined with the Euler characteristic test, minimal cuts allow us to derive hierarchical constraints on any given pair of singularities in an efficient way, without ever solving the Landau equations. We call this class of hierarchical constraints---which follow from identifying the minimal cuts that can give rise to a singularity, and which thus act as its earliest progenitors---\emph{genealogical constraints}.

\vspace{.1cm}
\noindent {\bf Two-Mass Easy Box Revisited}
\vspace{.1cm}

As a first demonstration of genealogical constraints, let us return to the two-mass easy box. The (unique) minimal cut associated with each of the Mandelstam variables is depicted in Figure~\ref{fig:two_mass_easy_box}. The minimal cut associated with each symbol letter can be constructed out of these by simply combining the cuts associated with each of the variables the letter depends on. For instance, the minimal cut for $s_{12} - s_{123}$ is given by cutting the lines associated with $\alpha_1$ and $\alpha_4$ (to resolve $s_{12}$), as well as $\alpha_2$ and $\alpha_4$ (to resolve $s_{123}$). Proceeding in this way, it is not hard to check that the minimal cuts for all of the singularities in Table~\ref{tab:two_mass_easy_box_constraints} exactly match the Feynman parameter contours that are pinched.

To derive genealogical constraints from these minimal cuts, we then see if the the Euler characteristic drops in a given kinematic limit, after dropping the $\alpha_i=0$ boundaries for each of the cut propagators. For instance, to check whether a discontinuity around $s_{45}=0$ can follow one around $s_{12}=0$, we compute $|\chi(Y_{1,4})|$ before and after setting \( s_{45} = 0 \). Since we find that it does not change, we conclude this sequence of discontinuities cannot appear. Conversely, the Euler characteristic \( |\chi(Y_{1,4})| \) does decrease when we set \( s_{12} - s_{123} = 0 \), indicating that a \( s_{12} - s_{123} \) discontinuity can still occur. Proceeding in this way, we reproduce the complete set of constraints we found before, as seen in Figure~\ref{fig:two_mass_easy_hierarchicial}.

\vspace{.1cm}
\noindent {\bf Genealogical Constraints at Two Loops}
\vspace{.1cm}

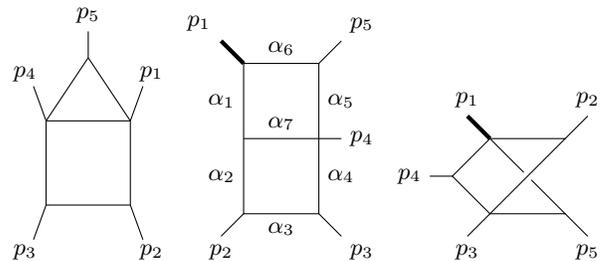
\begin{figure}
    \begin{tikzpicture}[scale=0.7]
        \coordinate (a) at (-.8,.8);
        \coordinate (b) at (.8,.8);
        \coordinate (c) at (0,2);
        \coordinate (d) at (.8,-.8);
        \coordinate (e) at (-.8,-.8);
        \draw[] (a) to (b);
        \draw[] (b) to (c);
        \draw[] (c) to (a);
        \draw[] (d) to (e);
        \draw[] (e) to (a);
        \draw[] (b) to (d);
        \draw[] (a) --++ (110:0.7);
        \draw[] (b) --++ (70:0.7);
        \draw[] (c) --++ (90:0.5);
        \draw[] (d) --++ (-70:0.7);
        \draw[] (e) --++ (-110:0.7);
        \node[] at (1.2,-1.7) {$p_2$};
        \node[] at (-1.2,-1.7) {$p_3$};
        \node[] at (-1.2,1.7) {$p_4$};
        \node[] at (1.2,1.7) {$p_1$};
        \node[] at (0,2.8) {$p_5$};        
    \end{tikzpicture}
\begin{tikzpicture}
    \draw (0,0) -- (1,0) node[midway, above] {$\alpha_6$} 
                -- (1,-1) node[midway, right] {$\alpha_5$}
                -- (1,-2) node[midway, right] {$\alpha_4$}
                -- (0,-2) node[midway, below] {$\alpha_3$}
                -- (0,-1) node[midway, left] {$\alpha_2$}
                -- cycle node[midway, left] {$\alpha_1$};
    \draw (1,-1) -- (0,-1)node[midway, above] {$\alpha_7$};
    \draw[line width=1.6pt] (0,0) -- (-0.3,0.3) node[above left] {$p_1$};
    \draw (1,0) -- (1.3,0.3) node[above right] {$p_5$};
    \draw (1,-1) -- (1.3,-1) node[right] {$p_4$};
    \draw (1,-2) -- (1.3,-2.3) node[below right] {$p_3$};
    \draw (0,-2) -- (-0.3,-2.3) node[below] {$p_2$};
\end{tikzpicture}
\begin{tikzpicture}
    \coordinate (n1) at (0,0);  
    \coordinate (n2) at (1,0);  
    \coordinate (n3) at (1,-1); 
    \coordinate (n4) at (0,-1); 
    \coordinate (n5) at (-0.5,-0.5);  
    \coordinate (n1p) at (.45,-.45);  
    \coordinate (n1pp) at (.55,-.55);  

    \coordinate (n6) at (-0.3,0.3); 
    \coordinate (n7) at (1.3,0.3);
    \coordinate (n8) at (1.3,-1.3);
    \coordinate (n9) at (-0.3,-1.3);
    \coordinate (n10) at (-0.8,-0.5);

    \draw (n1) -- (n2) -- (n4) -- (n5) -- (n1);
    \draw (n1pp) -- (n3) -- (n4);
    \draw (n1) -- (n1p);

    \draw[line width=1.6pt] (n1) -- (n6);
    \draw (n2)--(n7);
    \draw (n3)--(n8);
    \draw (n4)--(n9);
    \draw (n5)--(n10);

    \node[above] at (n6) {$p_1$};
    \node[above] at (n7) {$p_2$};
    \node[below] at (n8) {$p_5$};
    \node[below] at (n9) {$p_3$};
    \node[left] at (n10) {$p_4$};

\end{tikzpicture}
    \caption{The massless triangle-box diagram, and a pair of double box diagrams involving a single massive leg.}\label{fig:2loopgraphs}
\end{figure}

We now illustrate the power of these constraints at two loops, starting with the massless triangle-box in Figure~\ref{fig:2loopgraphs}. This integral was computed in~\cite{Chicherin:2018mue}, and involves 24 symbol letters. Similar to the two-mass easy box, each threshold singularity in this graph has a unique minimal cut, making it easy to construct the minimal cuts associated with more complicated singularities as before. We then find that we can derive 156 genealogical constraints; for instance, the letters in the set $\{s_{45} + s_{15}, s_{23} - s_{15}, s_{23} - s_{45}, s_{34} + s_{15}, s_{23} + s_{34} - s_{15}, s_{23} - s_{45} - s_{15}\}$ can never appear after the letter $s_{12}$. In addition, two of the algebraic letters that appear in this integral cannot follow $s_{12}$, as they give rise to logarithmic branch cuts that coincide with one of these restricted rational letters. We have explicitly confirmed that these constraints are satisfied through weight seven in dimensional regularization, and find that we miss only 31 further constraints of this type.

Let us now consider the pair of double-box diagrams in Figure~\ref{fig:2loopgraphs}. These integrals were computed in~\cite{Abreu:2020jxa,Abreu:2021smk}. In the planar example, some thresholds no longer have a unique minimal cut; for example, $p_1^2$ can be isolated by cutting either the lines associated with $\{\alpha_{1},\alpha_6\}$ or with $\{\alpha_2,\alpha_6,\alpha_7\}$. Thus, when determining the minimal cuts for singularities like $p_{1}^2-s_{234}$, we must consider all ways of combining the different cuts that isolate $p_{1}^2$ and $s_{234}$. New complications also arise in the non-planar case, as we now need to consider different ways of expressing each singularity. For instance, the singularity $p^2_{1} - s_{15} + s_{23} - s_{45}=0$ seemingly gives rise to a five-particle cut, but can be rewritten as $s_{14}=0$, which only has a three-particle cut. In this case, we can discard the former cut, since it involves cutting a superset of the propagators that are cut in the latter one. Proceeding in this way, we find 620 and 540 genealogical constraints on the symbol letters that appear in these planar and non-planar double-boxes, respectively, which we have have verified against the explicit results in~\cite{Abreu:2020jxa,Abreu:2021smk} (finding that we miss only 25 and 9 further empirical restrictions of this type). 
Like before, some of these restrictions apply to algebraic letters; however, we note that one has to be careful when multiple algebraic letters give rise to the same logarithmic branch cut (as happens here), since this branch cut cancels out in ratios of these letters. We include these two-loop predictions in ancillary files.

We can also probe how powerful these genealogical constraints are compared to the extended Steinmann relations, for instance in the example of the planar double box. A space of 679 integrable weight-four symbols that satisfy the first entry condition can be constructed out of the 37 symbol letters that appear in this integral (for more detail on how to construct such spaces, see for instance~\cite{Caron-Huot:2020bkp}). Imposing the extended Steinmann relations, which tell us that $s_{12}$ and $s_{15}$ cannot appear next to each other, reduces this number to 569, while imposing our genealogical constraints (which in this case also imply the extended Steinmann ones) brings it down further, to 264.\footnote{In fact, we observe that it is sufficient to impose the genealogical constraints that restrict which letters can appear after threshold discontinuities; integrability then seems to ensure that all other genealogical constraints are satisfied.} Thus, we see that these new constraints are significantly more restrictive than extended Steinmann.

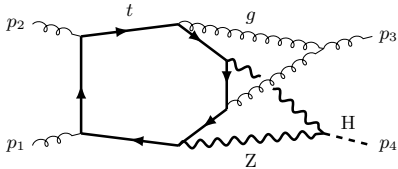
\begin{figure}
\centering
\resizebox{.3\textwidth}{!}{\begin{tikzpicture}[scale=2]
 
     \coordinate (n1) at (0,0); 
     \coordinate (n2) at (0,-0.8); 
     \coordinate (n3) at (0.8,0.1); 
     \coordinate (n4) at (1.2,-0.2); 
     \coordinate (n4a) at (1.48,-0.33);
     \coordinate (n4b) at (1.58,-0.43);
     \coordinate (n5) at (1.2,-0.6);  
     \coordinate (n6) at (0.8,-0.9);  
     \coordinate (n7) at (2,-0.1);  
     \coordinate (n8) at (2,-0.8);  

     \coordinate (k1) at (-0.4,0.1); 
     \coordinate (k2) at (-0.4,-0.9);
     \coordinate (k7) at (2.4,0);
     \coordinate (k8) at (2.4,-0.9);

     \draw [quark, line width=1.2pt](n2) -- (n1);
     \draw [quark, line width=1.2pt] (n1) -- (n3) node[midway, above=0.1cm] {$t$};
     \draw[quark, line width=1.2pt] (n3) -- (n4);
     \draw[quark, line width=1.2pt] (n4) -- (n5);
     \draw[quark, line width=1.2pt] (n5) -- (n6);
     \draw[quark, line width=1.2pt] (n6) -- (n2);
     \draw [gluon] (n3) -- (n7) node[midway, above=0.1cm] {$g$};
     \draw [gluon] (n5) -- (n7);
     \draw [dphoton,line width=1.2pt] (n4) -- (n4a);
     \draw [dphoton,line width=1.2pt] (n4b) -- (n8);
     \draw [dphoton,line width=1.2pt] (n6) -- (n8) node[midway, below=0.1cm] {${\rm Z}$};

     \draw [gluon] (k1) -- (n1);
     \draw [gluon] (k2) -- (n2);
     \draw [gluon] (n7) -- (k7);
     \draw [line width=1.2pt, dashed] (n8) -- (k8) node[midway, above=0.05cm] {${\rm H}$};

     \node[right] at (k8) {$p_4$};
     \node[right] at (k7) {$p_3$};
     \node[left] at (k1) {$p_2$};
     \node[left] at (k2) {$p_1$};

\end{tikzpicture}}
\caption{Example three-loop diagram that contributes to $gg \to Hg$, which involves a top quark loop and a pair of $Z$ boson propagators.}
\label{fig:gg_to_Hg}
\end{figure}

\vspace{.1cm}
\noindent {\bf Conclusions}
\vspace{.1cm}

In this letter, we have presented a practical method for deriving a powerful set of genealogical constraints on the discontinuities of any Feynman integral. In the two-loop examples we have studied, we have found that this method misses very few of the constraints that could follow from the hierarchical principle. Moreover, these predictions can be made efficiently (for instance, using the PLD code~\cite{Fevola:2023kaw, Fevola:2023fzn}), so it is not a stretch to derive constraints on complicated diagrams. For example, we find that the three-loop diagram in Figure~\ref{fig:gg_to_Hg} should satisfy
\begin{equation}
\Delta_{s_{12} + s_{23} - m_H^2 + m_Z^2} \cdots \Delta_{m_H^2 - 4 m_Z^2} \cdots I = 0  \, ,
\end{equation}
where $m_Z$ is the mass of the internal $Z$ bosons, and $m_H^2 = p_4^2$ is the mass of the external Higgs.

While we have restricted our attention to scalar integrals in this paper, we plan to describe how this method can be extended to integrals involving numerators and propagators raised to higher powers in an upcoming work~\cite{to_appear}. The same basic method can also be applied in momentum space, which may make it easier to connect with the types of numerators used by much of the community. 

Finally, in addition to the clear conceptual interest these hierarchical constraints hold, we expect them to provide extremely useful input for bootstrap methods. Since genealogical constraints can be derived in dimensional regularization for processes involving any masses, they are ideal for studying the integrals that contribute to important Standard Model processes.

\vspace{.1cm}
\noindent {\bf Acknowledgements}
\vspace{.1cm}

We thank Ruth Britto, Lance Dixon, Sebastian Mizera, Erik Panzer, Andrzej Pokraka, Matthew Schwartz, Simon Telen, and Cristian Vergu for stimulating discussion and fruitful collaboration on related work. This work has been supported by the Royal Society grant URF$\backslash$R1$\backslash$221233. H.S.H. gratefully acknowledges funding provided by the William D. Loughlin Membership, an endowed fund of the Institute for Advanced Study. This material is based upon work supported by the U.S. Department of Energy, Office of Science, Office of High Energy Physics under Award Number DE-SC0009988.

\bibliographystyle{apsrev4-1}
\bibliography{refs}

\end{document}